# Power Dependent Resonant Frequency of a Microwave Cavity due to Magnetic Levitation

N. K. Raut, J. Miller, H. Hart, R. Chiao, and J. E. Sharping

*Abstract*—Levitation of a magnet by superconductor has been an active area of research to explore the quantum mechanical phenomenon. One of the techniques used is to measure the levitation of a magnet placed inside the superconducting microwave cavity. The levitation height can be probed by measuring the change in microwave frequency. Here, we report measurements of the change in resonance frequency of the microwave cavity with the Meissner-levitated permanent magnet. The change in resonant frequency and quality factor was measured as a function of input power and temperature. The change in resonate frequency is likely due to the interaction of the magnet with the radio-frequency field inside the microwave cavity.

## I. INTRODUCTION

THE resonance frequency of a coaxial quarter-wave cavity can be approximated in terms of capacitance and inductance as [1]:

$$f = \frac{1}{2\pi\sqrt{LC}} \quad (1)$$

The capacitance term includes all the energy storing component of the cavity, whereas the inductance term includes all the loss mechanism [1]. The total inductance (L) can be divided into the geometry ($L_g$) and kinematic ($L_k$) inductance [2]. The $L_g$ value depends upon the shape of the cavity and is constant, whereas the $L_k$ term is dependent upon the external magnetic field as $L_k$ (B) ∝ $\lambda$ (B), where $\lambda$ is the penetration depth [3]. According to the two-fluid model, the density of the cooper pair ($n_s$) changes as a function of temperature as:

$$\frac{n_s}{n} = \left[1 - \left(\frac{T}{T_C}\right)^4\right], \quad (2)$$

where $n$ is the total number of electrons, and $T_C$ is the critical temperature of the superconductor [4]. The London penetration depth estimates the length up to which an external magnetic field penetrates [5]. It also depends on the density of the cooper pairs as:

$$\lambda_L = \sqrt{\frac{m_e c^2}{4\pi n_s e^2}} \quad (3)$$

Here, $m_e$ is the bare electron mass in the vacuum, $e$ is the electronic charge, and $c$ is the velocity of the light. The higher power that is put inside the superconducting cavity could reduce the density of the cooper pairs [6]. This in turn increases the penetration depth [7]. This phenomenon could help the levitated magnet change its position.

## II. EXPERIMENTAL SET UP

Figure 1 (a) shows a sketch of our experimental setup [8]. It consists of two crucial components: a microwave cavity and a magnet. A spherical neodymium magnet (N52) of radius 0.5 mm with strength 1.47 T is placed on the stub of the cavity similar to that proposed by Reagor et al. [9]. It has a long and wide outer cylinder of radius 7 mm and a height of 55 mm. Compared to the outer cylinder, the stub is 5 mm tall with radius of 2 mm. The stub established a quarter-wave resonance. One of the advantages of this design is that the higher order modes are separated by ∼ 2 GHz.

The cavity was fabricated from a single piece of bulk 6061 aluminum, a type-I superconductor of transition temperature, $T_c$ ∼ $1.2 K. The magnet is restricted on the stub by a snugly fitted plastic sleeve. The cavity-magnet system is fixed on the base plate of the dilution refrigerator. The fridge is cool-down in the automated way to its base temperature. In our experiment, the fridge is heated up and cooled down in a controlled way to the desired temperature and cooling rate. We have used a heater that is set in the mixing chamber to control the temperature. The temperature is adjusted by circulation of the $He_3$:$He_4$ mixture. For the data acquisition, a vector network analyzer (VNA) HP8720 is used. Full 2-port calibration is performed before each set of measurements using a special calibration kit.

The electric field of the cavity is localized around the edge of the stub (1.5 mm ≤ x ≤ 2 mm). Moreover, the electromagnetic field of the cavity decays exponentially ($e^{-\beta z}$) from the stub to the open end of the cavity, where $\beta$ is the propagation constant and $z$ is the vertical distance from the stub [9]. The high field

Corresponding author: N. K. Raut.
N. K. Raut was previously with with University of California, 5200 North Lake Road, 95343, Merced, CA, USA
N. K. Raut is currently with Thomas Jefferson National Laboratory, 600 Kelvin Dr, Suite 8, Newport News, VA 23606, USA (e-mail: raut@jlab.org).
J. Miller is with University of California, 5200 North Lake Road, 95343, Merced, CA, USA

H. Hart is with University of California, 5200 North Lake Road, 95343, Merced, CA, USA
R. Chiao is with University of California, 5200 North Lake Road, 95343, Merced, CA, USA
J. E. Sharping is with University of California, 5200 North Lake Road, 95343, Merced, CA, USA ABCD-123456789.

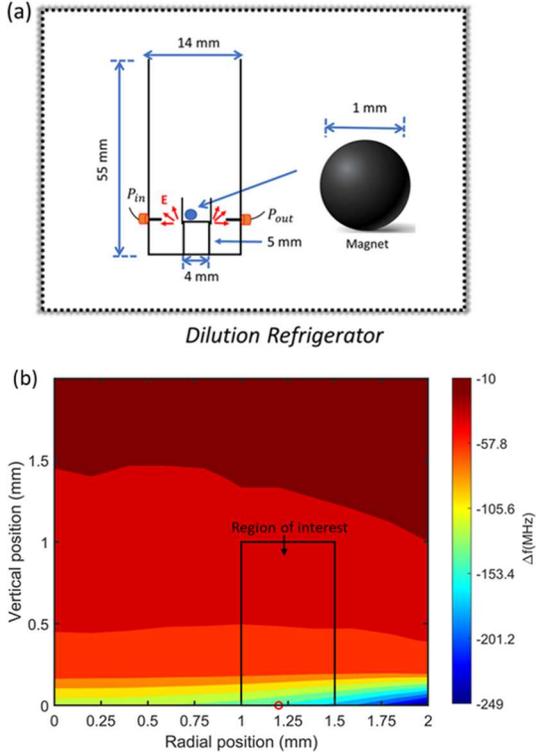

Fig. 1. (a) Schematic of our experimental set up. A spherical magnet of radius 0.5 mm is placed on the surface of the stub of the cavity. The cavity-magnet is then set up on the base plate of the dilution refrigerator. (b) FE simulations on probing resonance frequency of the cavity-magnet system. The effect is measured based on the change in the frequency [16].

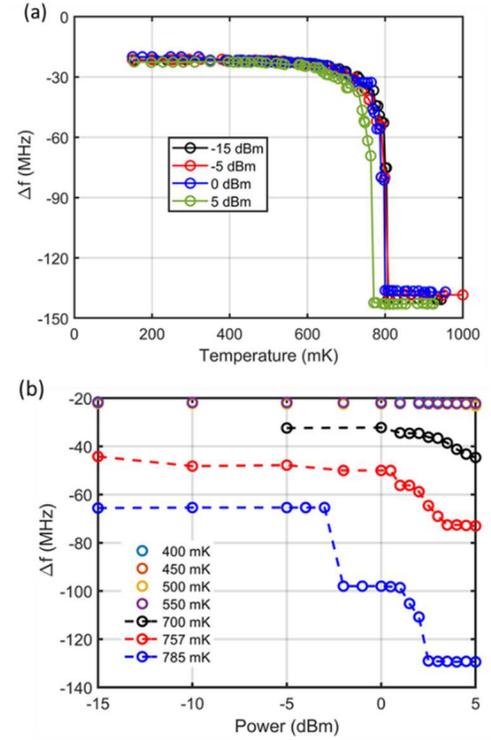

Fig. 2. (a) Resonance frequency of the cavity as a function of temperature set at different powers, (b) the temperature of the cavity is held fixed and the power is ramped up from -15 dBm to 5 dBm.

concentration at a small area is extremely sensitive to the external perturbation. Such interaction results into a large change in the resonance frequency of the cavity [10]. In this work, a magnet is levitated from the edge of the stub. Here, the amount of interaction between the magnet and localized field determines the amount of frequency shift.

To better understand the frequency shift, finite element (FE) simulations was done. The change in the frequency with respect to the position of magnet is shown in Fig. 1(b). The difference in frequency is calculated from the bare cavity frequency. In the simulations, the magnet is manually shifted from the center (x, z= 0 mm, 0 mm) to the edge (x, z= 0 mm, 2 mm) of the stub. Same procedure is repeated to the distance of 2 mm above the stub. The region enclosed by the black lines is the region of our interest. We expect magnetic levitation in this region [11].

There are two unique frequency shift patterns due to the radial and vertical displacements of the magnet. When the magnet moves radially, its interaction with the concentrated electric field increases. This type of magnetic displacement cause downshift in the frequency. However, the trend of the frequency shift is opposite for magnetic levitation. The lift of the magnet induces upward shift in the frequency. In both the cases, the amount of frequency change is highest at the edge of the stub [12].

## III. CRYOGENIC MEASUREMENTS

### A. Effect of the Input Power

The cavity with a magnet resting on the stub is first let automated cooldown to its base temperature ( ~135 mK). Then, it is slowly warmed up to 1 K using the heater attached to the mixing chamber. Same procedure is repeated for different input powers in test. The change in frequency as a function of temperature at different input power level is shown in Fig. 2(a). The frequency shows the sudden jumps at higher temperature indicating the large change in penetration depth due to the change in effective inductance of the cavity-magnet system. The largest downshift is due to the phase transition of the superconductor into the normal state.

Figure 2 (a) shows the change in resonance frequency as a function of temperature at different power level. The three low powers (-15 dBm, -5 dBm and 0 dBm) caused the phase transition around 800 mK. However, the significant change in the transition temperature is seen at the input power of 5 dBm. The 5 dBm power induced the phase transition at lower temperature. A detail study of the effect of the input power on the resonance frequency of the cavity shown in Fig. 2(b). The temperature of the cavity was held fixed and the input power to the cavity was ramped up from -15 dBm to 5 dBm. Each data is taken for period of one minute. At lower temperatures (< 550 mK), there was a small or negligible change in the resonance frequency of the cavity due to the power ramp up. The observation could be either due to the robust cooling such that the heating due to the power is minimal or the resolution of instrument is limited. More studies are needed to confirm these results. At the higher temperatures (>700 mK), the increase in power induces at least 10 MHz of frequency shift. The effect is prominent at 757 mK and 785 mK.

is likely due to the robustness of the cavity-magnet system. It is to be noted that this switching in frequency was not observed at 785 mK, where the change in quality factor was observed due to the power switching as shown in Fig.3.

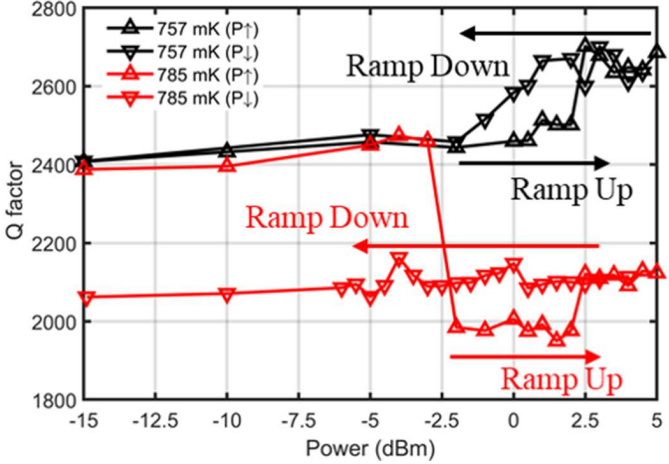

Fig. 3. The total quality factor of the cavity-magnet system as a function of input power at different temperature.

### B. Quality Factor as a Function of Power

We have seen significant effect of the input power on the resonance frequency of the cavity at 757 mK and 785 mK. To further investigate this effect, quality factor as a function of power was measured. The total quality factor of the cavity quantifies the total amount of energy that the cavity can store compared to the total losses in one radio frequency (RF) cycle [13].

Figure 3 shows the total Q of the cavity-magnet system as a function of the power at 757 mK and 785 mK. The power is ramped up from -15 dBm to 5 dBm and is ramped back down to -15 dBm. The data acquisition is done every minute. At 757 mK, improvement in the Q is seen during the increase of power. The trend of the Q was sustained during the power ramp-down as well [14]. However, at 785 mK a sharp drop in Q is observed around -2.5 dBm. There was no recovery of Q after that. We conclude the superconductor is robust enough against the external magnetic field at 757 mK. However, the input power has a destructive effect on Q at 785 mK. The hysteresis in Q likely due to the trapped magnetic field during the power ramp down [15].

### C. Power variation at 757 mK

As shown above, the cavity-magnet system shows its robustness at 757 mK. At this temperature, we have measured the change in frequency at different power level. The data were recorded for 5 minutes of duration. This allows us to thermally stabilized the cavity before the measurement were performed. As shown in Fig. 4(a), the large change in frequency was observed for input power ≥5dBm. The second set of measurement were done with switching the power from high to low and and repeating the measurements at different power level as shown in Fig. 4(b). The switching feature in frequency

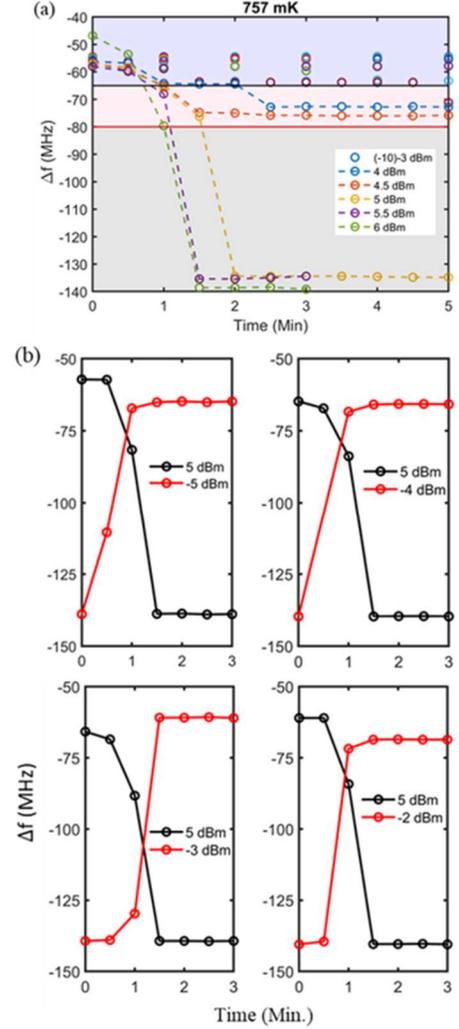

Fig. 4. (a) The change is frequency at different power level measured at 757 mK (b) change in frequency during power switch from 5 dBm to (-5, -4, -3, & -2) dBm respectively.

### IV. CONCLUSION

We have probed the resonance frequency of the superconducting quarter-wave cavity with the variation of the input power. The effect is first checked by slowly heating the cavity at powers -15 dBm, -5 dBm, 0 dBm, and 5 dBm from the base temperature of the fridge (135 mK) to 1 K. In this test, the higher power (5 dBm) induced superconducting phase transition at significantly lower temperature. Furthermore, measurement was done by keeping temperature of the cavity fix at different temperature. The cavity magnet system showed the interesting switching in frequency and quality factor depending on the input power and temperature. The set up could be useful to understand the quantum mechanical phenomenon of cavity-magnet coupled system.

ACKNOWLEDGMENT

The author would like to acknowledge Dr. Pashupati Dhakal for useful discussion. Also, like to acknowledge Dr. Robert Rimmer for his support.